# Optical Tuning of Exciton and Trion Emissions in Monolayer Phosphorene

Jiong Yang,[1,†] Renjing Xu,[1,†] Jiajie Pei, [1,4,†] Ye Win Myint,[1] Fan Wang,[2] Zhu Wang,[3] Shuang Zhang,[1] Zongfu Yu,[3] and Yuerui Lu[1*]

[1]Research School of Engineering, College of Engineering and Computer Science, the Australian National University, Canberra, ACT, 0200, Australia

[2]Department of Electronic Materials Engineering, Research School of Physics and Engineering, the Australian National University, Canberra, ACT, 0200, Australia

[3]Department of Electrical and Computer Engineering, University of Wisconsin, Madison, Wisconsin 53706, USA

[4]School of Mechanical Engineering, Beijing Institute of Technology, Beijing, 100081, China

[†] These authors contributed equally to this work

[*] To whom correspondence should be addressed: Yuerui Lu (yuerui.lu@anu.edu.au)

**Abstract:**

**Monolayer phosphorene provides a unique two-dimensional (2D) platform to investigate the fundamental dynamics of excitons and trions (charged excitons) in reduced dimensions. However, owing to its high instability, unambiguous identification of monolayer phosphorene has been elusive. Consequently, many important fundamental properties, such as exciton dynamics, remain underexplored. We report a rapid, noninvasive, and highly accurate approach based on optical interferometry to determine the layer number of phosphorene, and confirm the results with reliable photoluminescence measurements. Furthermore, we successfully probed the dynamics of excitons and trions in monolayer phosphorene by controlling the photo-carrier injection in a relatively low excitation power range. Based on our measured optical gap and the previously measured electronic energy gap, we determined the exciton binding energy to**

be ~0.3 eV for the monolayer phosphorene on SiO$_2$/Si substrate, which agrees well with theoretical predictions. A huge trion binding energy of ~100 meV was first observed in monolayer phosphorene, which is around five times higher than that in transition metal dichalcogenide (TMD) monolayer semiconductor, such as MoS$_2$. The carrier lifetime of exciton emission in monolayer phosphorene was measured to be ~220 ps, which is comparable to those in other 2D TMD semiconductors. Our results open new avenues for exploring fundamental phenomena and novel optoelectronic applications using monolayer phosphorene.

**Keywords**: Monolayer phosphorene, exciton, trion, optical injection

**Introduction**

Phosphorene is a recently developed two-dimensional (2D) material that has attracted tremendous attention owing to its unique anisotropic manner[1-6], layer-dependent direct band gaps[7,8], and quasi-one-dimensional (1D) excitonic nature[9,10], which are all in drastic contrast with the properties of other 2D materials, such as graphene[11] and transition metal dichalcogenide (TMD) semiconductors[12-14]. Monolayer phosphorene has been of particular interest in exploring technological applications and investigating fundamental phenomena, such as 2D quantum confinement and many-body interactions[9,15]. However, such unique 2D materials are unstable in ambient conditions and degrade quickly[8,16]. Particularly, monolayer phosphorene is expected to be much less stable than few-layer phosphorene[16], hence making its identification and characterization extremely challenging. There is a huge controversy on the identification of very few-layer (one or two layers) phosphorene and thus on their properties[16-18]. This controversy was primarily due to the lack of a robust experimental technique to precisely identify the monolayer phosphorene. Consequently, many important fundamental properties of monolayer phosphorene, such as its excitonic nature, remain elusive.

In this study, we propose and implement a rapid, noninvasive, and highly accurate approach to determine the layer number of mono- and few-layer phosphorene by using optical interferometry. The identification is clearly confirmed by the strongly layer-dependent peak energies in the measured photoluminescence (PL) spectra. More importantly, we successfully probed the exciton and trion dynamics in monolayer phosphorene by controlling the photo-carrier injection at a very low excitation power range. The exciton binding energy of monolayer phosphorene on a $SiO_2$/Si substrate was determined to be ~0.3 eV; this result agrees well with the theoretical prediction that substrate screening strongly affects the exciton binding energy in monolayer phosphorene[15]. Furthermore, a high trion binding energy of ~100 meV (upper bound) was observed in the monolayer phosphorene on a $SiO_2$/Si substrate, which, again, agrees well with our theoretical calculation. In addition, time-resolved PL was used to characterize the critical carrier dynamics in monolayer phosphorene. The carrier lifetime of exciton emission in monolayer phosphorene was measured to be ~220 ps, a value comparable to those in other TMD semiconductors. Our results provide a new platform for the investigation of fundamental many-body interactions and to explore new optoelectronic applications using monolayer phosphorene.

**Materials and Methods**

*Sample preparation and characterization*. Mono- and few-layer phosphorene, graphene, and TMD semiconductor samples were mechanically exfoliated from bulk crystals and drily transferred onto $SiO_2$/Si (275 nm thermal oxide) substrates. A phase shifting interferometer (Vecco NT9100) was used to obtain all the OPL values for phosphorene samples. Monolayer phosphorene samples were put into a Linkam THMS 600 chamber and the temperature was set as −10 °C during the power-dependent PL and time-resolved PL (TRPL) measurements. The power-dependent PL and TRPL measurements were conducted in a setup which incorporates µ-PL spectroscopy and a time-correlated single photon counting (TCSPC) system. A linearly polarized pulse laser (frequency doubled to 522 nm, with 300 fs pulse width and 20.8 MHz repetition rate) was directed to a high numerical aperture (NA = 0.7) objective (Nikon S Plan 60x). PL signal was collected by a grating spectrometer, thereby either recording the PL spectrum through a charge coupled device (CCD) (Princeton Instruments, PIXIS) or detecting the PL intensity decay by a Si single photon avalanche diode (SPAD) and the TCSPC (PicoHarp 300) system with a resolution of ~40 ps.

For few-layer phosphorene (2L to 5L), the PL measurements were conducted using a T64000 micro-Raman system equipped with a InGaAs detector, along with a 532 nm Nd:YAG laser as the excitation source. For all the PL measurements for 2-5L phosphorene samples, the sample was placed into a microscope-compatible chamber with a slow flow of protection nitrogen gas to prevent sample degradation at room temperature. To avoid laser-induced sample damage, all PL spectra from two- to five-layer phosphorene were recorded at low power level of P ~20 µW.

*Numerical Simulation.* Stanford Stratified Structure Solver (S4)[19] was used to calculate the phase delay. The method numerically solves Maxwell's equations in multiple layers of structured materials by expanding the field in the Fourier-space.

**Results and Discussion**

Both atomic force microscopy (AFM) and Raman spectroscopy have been used to reliably determine the sample thickness of TMD semiconductors with monolayer precision[20]. However, these two methods are not reliable for the identification of very-few-layer phosphorene (one or two layers). The scanning rate of AFM is slow compared to the fast degradation of very-few-layer phosphorene in ambient conditions and AFM can easily generate an error of one or even two layers, owing to the large surface roughness in very-few-layer phosphorene samples. AFM can also introduce potential contaminants that might affect further characterizations on the same sample. Unlike in TMD semiconductors, where Raman mode frequency has a monotonic dependence on the layer number, phosphorene has a non-monotonic dependence owing to the complicated Davydov-related effects[18]. Moreover, the relatively high-power laser used in Raman spectroscopy can significantly damage the phosphorene samples.

To overcome the aforementioned challenges, we propose and implement a rapid, noninvasive, and highly accurate approach to determine the layer number by using optical interferometry (Figure 1). Specifically, we measure the optical path length (OPL) of the light reflected from the phosphorene that was mechanically exfoliated onto a SiO$_2$/Si substrate (275 nm thermal oxide). The OPL is determined from the relation: $OPL_{BP} = -\frac{\lambda}{2\pi}(\phi_{BP} - \phi_{SiO_2})$, where λ is the wavelength of the light source and is equal to 535 nm, and $\phi_{BP}$ and $\phi_{SiO_2}$ are the phase shifts of the light reflected from the phosphorene flake and the SiO$_2$/Si substrate (Figure 1c inset), respectively. The direct relationship between the OPL and the layer number is firmly established by a first-principle calculation and experimental calibration, as shown in Figure 1d. Even though the thickness of monolayer phosphorene is less than 1 nm, its OPL is significantly larger than 20 nm owing to the multiple interfacial light reflections (Supplementary Information). That is, the virtual thickness of a phosphorene flake is amplified by more than 20 times in the optical interferometry, making the flakes easily identifiable. In the experiment, phase-shifting interferometry (PSI) is used to measure the OPL by analyzing the digitized interference pattern. In contrast to the highly focused and relatively high-power laser used in Raman system, PSI uses almost non-focused and very low-density light from a light-emitting diode (LED) source to achieve fast imaging (Supplementary Information), which inflicts no damage to the phosphorene samples. The step change of the OPL is ~20 nm for each additional phosphorene layer, as indicated by the red dots in Figure 1d. Considering that the accuracy of the instrument is ~0.1 nm, a step change of 20 nm yields extremely robust identification of the layer number. Statistical OPL values for phosphorene from mono- to six-layer (1L to 6L) were collected and at least five different samples were measured with the PSI system for each layer number. The measured OPL values agree very well with our theoretical calculations (Figure 1d). Recently, we also successfully used PSI to quickly and precisely identify the layer numbers of TMD atomically thin semiconductors[21].

Subsequent to PSI measurement, the sample was placed into a Linkam THMS 600 chamber, at a temperature of $-10\ ^\circ C$ with a slow flow of nitrogen gas to prevent degradation of the sample[8]. The low temperature ($-10\ ^\circ C$) is a very crucial factor because it can freeze the moisture in the chamber and significantly delay the sample degradation. Under $-10\ ^\circ C$ and nitrogen protection, monolayer phosphorene samples can survive for several hours in the chamber. However, even in a temperature of $-10\ ^\circ C$ with nitrogen gas protection, the monolayer phosphorene sample was found damaged when the power of the pulsed laser was higher than 1.15 µW (Figure S3). When the chamber temperature was raised from $-10\ ^\circ C$ to room temperature, the monolayer phosphorene was oxidized immediately and the PL signal disappeared. In contrast to monolayer phosphorene, 2L and 3L phosphorene samples can survive for more than fifteen hours under $-10\ ^\circ C$ and for several hours when the chamber temperature was raised to room temperature.

Because of the strongly layer-dependent peak energies and the direct band gap nature of phosphorene, we are able to further confirm the layer number identification by measuring their corresponding peak energies of the PL emission (Figure 2). Figure 2a shows the normalized PL spectra of the mono- to five-layer phosphorene samples. The emission peak of the PL spectrum for monolayer phosphorene is at 711 nm, corresponding to a peak energy of 1.75 eV. This PL peak energy value was measured at $-10\ ^\circ C$ and it is expected not to vary too much at room temperature. Temperature-dependent PL measurements were conducted on 2L and 3L phosphorene samples from 20 °C down to −70 °C; very minor shifts of -0.112 meV/°C and -0.032 meV/ °C with temperature were observed for 2L and 3L phosphorene samples, respectively (Figure S4). Assuming a similar low temperature dependence for monolayer phosphorene, its PL peak energy at room temperature would be only ~1–4 meV lower than the measured value at −10 °C. Combining the results of our previous work[8] on few-layer phosphorene (2L to 5L) with the results obtained from our recent samples (1L to 5L), it can be clearly observed that the peak energy of PL emission shows unambiguous layer dependence (Figure 2b). For each layer number, at least three samples were characterized; the measured peak energies for 1L to 5L phosphorene are 1.75 ± 0.04, 1.29 ± 0.03, 0.97 ± 0.02, 0.84 ± 0.02, and 0.80 ± 0.02 eV, respectively. The PL emission energy of the phosphorene samples with higher layer number (> 5) is beyond the measurement wavelength range (up to 1600 nm) of our system.

The peak energy of PL emission, also termed as optical gap ($E_{opt}$), is the difference between the electronic band gap ($E_g$) and the exciton binding energy ($E_b$) (Figure 2b inset). Owing to the strong quantum confinement effect, free-standing monolayer phosphorene is expected to have a large exciton binding energy of ~0.8 eV[9, 15], whereas this value is expected to be only ~0.3 eV for monolayer phosphorene on a $SiO_2$/Si substrate because of the increased screening from the substrate[15]. If we use the measured electronic energy gap of ~2.05 eV by Pan *et al* [22] using scanning tunneling spectroscopy and our measured optical gap of 1.75 eV in monolayer phosphorene, the exciton binding energy of monolayer phosphorene on $SiO_2$/Si substrate is determined to be ~0.3 eV, which agrees very well with the prediction[15]. The optical gaps in phosphorene increase rapidly with decreasing layer number because of the strong quantum confinement effect and the van der Waals interactions between the neighboring sheets in few-layer phosphorene[9, 23]. We used a power law form to fit the experimental data and obtained the fitting curve of $E_{opt} = \frac{1.486}{N^{0.686}} + 0.295$, where $E_{opt}$ is the optical gap in unit of eV and $N$ is the layer number (Figure 2b). The layer-dependent optical gaps, as indicated in Figure 2b, agree very well with the theoretical predictions[7, 9]. For bulk phosphorene sample with large $N$ value, its optical gap approaches the limit value of ~0.295 eV, which matches very well with the measured energy gap (~0.3 eV) of bulk phosphorene[5, 24]. Previously, Ye *et al*[1] observed bright exciton PL emission at ~1.45 eV from a monolayer phosphorene that is coated with a protection layer of PMMA. This protection layer might introduce some defect states to the sample, which could change the PL emission energy. Our samples have no any protection layers and we did not use any chemical treatment processes, which provide very clean surfaces in our samples for exciton nature investigations.

The dynamics of excitons and trions have been of considerable interest for fundamental studies of many-body interactions[13, 14], such as carrier multiplication and Wigner crystallization[25]. Monolayer phosphorene, whose excitons are predicted to be confined in a quasi-one-dimensional (1D) space[7, 9], provides an ideal platform for investigating remarkable exciton and trion dynamics in a reduced dimension. Here, we successfully probed the exciton and trion dynamics in monolayer phosphorene by controlling the photo-carrier injection in power-dependent PL measurements (Figure 3). The measured PL spectra exhibit two clear peaks with central wavelengths at ~705 nm (labeled as "A") and ~760 nm (labeled as "X"), whose intensities and peak positions are highly dependent on the excitation power (Figure 3a). We used Lorentzian curves to fit the measured PL spectra and extracted the spectral components of these two peaks, indicated by red and blue curves in Figure 3a. The intensity of higher-energy peak A increases almost linearly with laser power (Figure 3b), while that of lower-energy peak X gradually saturates at a relatively high photo-carrier injection. When the excitation power changed from 0.19 to 1.15 μW (Figure 3c), the optical gap of A emission ($E_A$) monotonically decreased by ~15 meV, while that of X emission ($E_X$) increased by ~35 meV.

The higher-energy peak A is attributed to exciton emission. The measured full width at half maximum (FWHM) of peak A decreases with increase in excitation power (Figure S5a), which suggests an increase in the exciton binding energy, and thus, a reduction in $E_A$ with the increase in power within the range of our experiments[26, 27]. Owing to the appropriate protection for the phosphorene, we obtained very stable PL spectra data and the measurement errors of the exciton and trion energies are minimized to be ~5 meV (Supplementary Information, Table S1).

The origin of peak X is particularly interesting. We believe the experimental observations do not support that peak X comes from the emission of localized excitons[28, 29]. First, the energy difference between the PL emissions from the localized and free excitons is not sensitive to the relatively low density of photo-carrier injection[28, 29]. But in our experiment, this energy difference decreases significantly from ~150 meV to ~100 meV when the power of excitation laser increases from 0.19 to 1.15 µW. Furthermore, localized excitons are expected to have a longer lifetime than free excitons because of the localization effect of squeezing an exciton in the zero-dimension-like state[28, 30]. However, our time-resolved PL measurements revealed that the carrier lifetime from X state is less than that from A state (discussed later). On the other hand, the origin of X state can be attributed to trions[13, 14]. The trion peak intensity saturates at high photo-carrier injection (Figure 3b) because the free carriers coming from initial doping are almost depleted during the formation of trions. The trion density estimated from our optical injection is comparable to the estimated initial doping level of our monolayer phosphorene sample (Supplementary Information).

The difference between the peak energies of exciton and trion PL emission is predicted to be[13, 31]:

$$E_A - E_X = E_{Tb} + \Delta E \tag{1}$$

where $E_{Tb}$ is the binding energy of trions and $\Delta E$ is the average energy needed to add one carrier into the free-carrier system. The energy difference $E_A - E_X$ represents the minimum energy for the removal of one electron (hole) from a negative (positive) trion. To convert a trion to an exciton, one of the two electrons (holes) in the negative (positive) trion will be unbound first ($+E_{Tb}$) and then added into the system ($+\Delta E$)[13, 31]. During PL excitation, the same amount of electrons and holes will be injected into the conduction band and valence band, respectively. At low excitation power (power close to zero), the system is in an equilibrium state and $\Delta E$ is approximately equal to the Fermi energy $E_F$ that is determined at the initial doping level[13, 31]. At higher PL excitation power with more carrier injection, the initial doping becomes less important because of the enhanced screening effect from the injected photo-carriers. In most conventional PL measurements with very high excitation power[10, 14, 32], $\Delta E$ can be neglected and the trion binding energy can be approximately taken as the energy difference $E_A - E_X$. Here, we used a very low power range as the PL excitation and were able to observe the unique effect of $\Delta E$. We started with a low power excitation at 0.19 μW, which has an injected photo-carrier concentration comparable to the initial doping level. As we increased the excitation power, the energy difference $E_A - E_X$ decreased, indicating a reduction of $\Delta E$. At high excitation power (1.15 μW), the initial doping becomes less important and $\Delta E$ can be approximately neglected; the energy difference $E_A - E_X$ of ~100 meV at high excitation power (1.15 μW) can be taken to be the upper limit of the trion binding energy in monolayer phosphorene. This value agrees well with the estimated trion binding energy in monolayer phosphorene of $E_{Tb} \sim 0.3\ E_b$[10, 23] and the calculated exciton binding energy in monolayer phosphorene on SiO$_2$/Si substrate is $E_b \sim 0.3$ eV[9]. In monolayer phosphorene, both positive and negative trions are expected to have similar binding energies, since the effective masses of electrons and holes are almost equal[7, 9].

Carrier lifetime has been considered to be a very critical parameter in semiconducting materials[33-35]. The accurate probing of the carrier lifetime in monolayer phosphorene can greatly help us in understanding its highly excitonic nature and to better explore its optoelectronic applications. In order to understand the origin of X emission peak, we used time-resolved PL to characterize the carrier lifetimes of these two different states, A and X. For the first time, we successfully measured the carrier lifetime of the A peak emission in monolayer phosphorene at 220±8 ps. The carrier lifetime of the X peak emission was found to be lower than the resolution of our system (~40 ps), which confirms the aforementioned reasoning that X peak should not be from the emission of localized excitons. In our experiments, a linearly polarized pulse laser (frequency doubled to 522 nm, with 300 fs pulse width and 20.8 MHz repetition rate) was used to excite the monolayer phosphorene samples. PL signal was collected by a grating spectrometer and the PL intensity decay was detected using a Si single-photon avalanche diode (SPAD) and the time-correlated single photon counting TCSPC (PicoHarp 300) system. The measured PL decay of the exciton peak of our monolayer phosphorene sample with a laser power of 1.15 μW is presented in Figure 4a with the system response to the excitation laser as reference. We fitted the measured decay curve with equation $I_{BP} = A\exp\left(-\frac{t}{\tau}\right) + B$, where $I_{BP}$ is the PL intensity, $A$ and $B$ are two constants, $t$ is time, and $\tau$ is carrier lifetime. Through fitting, the carrier lifetime $\tau$ was determined to be 211 ps. Power-dependent PL decay was conducted with pulse laser powers 0.19, 0.38, 0.77, and 1.15 μW, and the same fitting process was performed to obtain the carrier lifetimes to be 222, 219, 232, and 211 ps, respectively, showing no significant power dependence. The power dependent carrier lifetime measurement results of the monolayer phosphorene sample differ from the results of other TMD semiconductors, which show a decreasing carrier lifetime with increase in excitation laser power[35]. This independence of carrier lifetime of our monolayer phosphorene samples suggests that higher order processes such as exciton-exciton annihilation were negligible for the excitation power employed in our experiments[34]. We also compared the carrier lifetime data from monolayer phosphorene samples with those from other monolayer TMD semiconductors measured with the same system, as indicated in Figure 4b. The carrier lifetime values for monolayer phosphorene, $MoS_2$, $WSe_2$, and $MoSe_2$ were measured to be 221 ± 8, 277 ± 13, 448 ± 60, and 604 ± 43 ps, respectively.

**Conclusions**

In conclusion, we report a rapid, noninvasive, and highly accurate approach to determine the layer number of mono- and few-layer phosphorene using PSI. The identification is further confirmed by reliable, highly layer-dependent PL peak energies. These two methods provide definite references for future mono- and few-layer phosphorene layer number identification. The dynamics of excitons and trions in monolayer phosphorene was successfully characterized by controlling the photo-carrier injection in a relatively low excitation power range. Based on the measured optical gap and previously measured electronic energy gap, we determined the exciton binding energy to be ~0.3 eV for the monolayer phosphorene on $SiO_2$/Si substrate, which agrees well with theoretical predictions. A huge trion binding energy of ~100 meV (upper limit) was first observed in monolayer phosphorene on $SiO_2$/Si substrate. In addition, a carrier lifetime of 220 ps for the monolayer phosphorene was first measured, which is comparable to other 2D TMD semiconductors. Our results open new routes for both the investigation of 2D quantum limit in reduced dimensions and development of novel optoelectronic devices.

**Acknowledgements**

We wish to acknowledge support from the ACT node of the Australian National Fabrication Facility (ANFF). We also thank Professor Chennupati Jagadish, Professor Lan Fu, and

Professor Barry Luther-Davies from the Australian National University (ANU) for facility support. We acknowledge financial support from the ANU PhD scholarship, the China Research Council PhD scholarship, the National Science Foundation (USA) (grant number ECCS-1405201), the Australian Research Council (grant number DE140100805), and the ANU Major Equipment Committee.

**Figure caption list**

**Figure 1 | Robust identification of mono- and few-layer phosphorene by phase shifting interferometry (PSI). a,** Optical microscope image of a monolayer phosphorene (labeled as "1L"). Inset is the schematic of single-layer phosphorene molecular structure. **b,** PSI image of the dash line box area indicated in (a). **c,** PSI measured optical path length (OPL) values along the dash line indicated in (b). Inset is the schematic plot showing the PSI measured phase shifts of the reflected light from the phosphorene flake ($\phi_{BP}$) and the SiO$_2$/Si substrate ($\phi_{SiO_2}$). **d,** OPL values from simulation and experiment PSI measurements for phosphorene samples from 1L to 6L. For each layer number of phosphorene, at least five different samples were characterized for the statistical measurements. The red dash line is the linear trend for statistical data measured with the PSI system.

**Figure 2 | Photoluminescence (PL) characteristics of mono- and few-layer phosphorene. a,** PL spectra of the mono- to five-layer phosphorene samples. Each PL spectra is normalized to its peak intensity and system background. **b,** Evolution of PL peak energy with layer number of phosphorene from experimental PL spectra, showing a rapid increase in peak energy as the layer number decreases. The solid grey line is the fitting curve $E_{opt} = \frac{1.486}{N^{0.686}} + 0.295$, where $E_{opt}$ is the optical gap in unit of eV and $N$ is the layer number. Note: the PL for the monolayer phosphorene was measured at $-10$ °C, while others were measured at room temperature. For monolayer phosphorene, the difference between PL peak energies at $-10$ °C and at room temperature is estimated to be less than 5 meV. Inset: schematic energy diagram showing the electronic band gap ($E_g$), and the exciton binding energy ($E_b$).

**Figure 3 | Exciton and trion dynamics in monolayer phosphorene. a,** Measured PL spectra (solid grey lines) under various excitation laser power. PL spectra are fit to Lorentzians (solid red lines are the exciton components, solid blue lines are the trion components, and solid pink

lines are the cumulative fitting results). **b,** PL intensity of exciton (A) and trion (X) as a function of laser power. **c,** PL peak energy of exciton (A) and trion (X) as a function of laser power.

**Figure 4 | Time-resolved photoluminescence (TRPL) characteristics of monolayer phosphorene. a,** PL decay profile of the exciton peak in monolayer phosphorene, by using a pulse 522 nm (frequency doubled) excitation laser at a laser power of 1.15 μW and a time-correlated single photon counting system (TCSPC). A lifetime of 211 ps can be extracted from the PL decay curve by exponential fitting. The blue solid line curve is the exponential fitting of the experimental data and the grey dash line curve is the instrument response to the excitation laser pulse. Inset shows the power-dependent carrier lifetime of monolayer phosphorene at pump laser power of 0.19, 0.38, 0.77 and 1.15 μW. For different excitation laser power, the carrier lifetime of monolayer phosphorene stays ~ 220 ps and shows no significant power dependence. **b,** Carrier lifetime for excitons of monolayer phosphorene, $MoS_2$, $WSe_2$ and $MoSe_2$ semiconductors. All data were measured with a pulse 522 nm excitation laser and with a laser power of 1.15 μW.

# References

1. H. Liu, A. T. Neal, Z. Zhu, Z. Luo, X. Xu, D. Tománek *et al.* Phosphorene: An unexplored 2D semiconductor with a high hole mobility. *ACS Nano* 2014; **8**; 4033-4041.
2. M. Buscema, D. J. Groenendijk, S. I. Blanter, G. A. Steele, H. S. J. van der Zant, A. Castellanos-Gomez. Fast and broadband photoresponse of few-layer black phosphorus field-effect transistors. *Nano Lett* 2014; **14**; 3347-3352.
3. R. Fei, L. Yang. Strain-engineering the anisotropic electrical conductance of few-layer black phosphorus. *Nano Lett* 2014; **14**; 2884-2889.
4. F. Xia, H. Wang, Y. Jia. Rediscovering black phosphorus as an anisotropic layered material for optoelectronics and electronics. *Nat Commun* 2014; **5**.
5. L. Li, Y. Yu, G. J. Ye, Q. Ge, X. Ou, H. Wu *et al.* Black phosphorus field-effect transistors. *Nat Nanotechnol* 2014; **9**; 372-377.
6. T. Hong, B. Chamlagain, W. Lin, H.-J. Chuang, M. Pan, Z. Zhou *et al.* Polarized photocurrent response in black phosphorus field-effect transistors. *Nanoscale* 2014; **6**; 8978-8983.
7. J. Qiao, X. Kong, Z.-X. Hu, F. Yang, W. Ji. High-mobility transport anisotropy and linear dichroism in few-layer black phosphorus. *Nat Commun* 2014; **5**.
8. S. Zhang, J. Yang, R. Xu, F. Wang, W. Li, M. Ghufran *et al.* Extraordinary photoluminescence and strong temperature/angle-dependent raman responses in few-layer phosphorene. *ACS Nano* 2014; **8**; 9590-9596.
9. V. Tran, R. Soklaski, Y. Liang, L. Yang. Layer-controlled band gap and anisotropic excitons in few-layer black phosphorus. *Phys Rev B* 2014; **89**; 235319.
10. S. Zhang, R. Xu, F. Wang, J. Yang, Z. Wang, J. Pei *et al.* Extraordinarily bound quasi-one-dimensional trions in two-dimensional phosphorene atomic semiconductors. *arXiv:1411.6124* 2014.
11. A. K. Geim, K. S. Novoselov. The rise of graphene. *Nat Mater* 2007; **6**; 183-191.
12. B. Radisavljevic, A. Radenovic, J. Brivio, V. Giacometti, A. Kis. Single-layer mos2 transistors. *Nat Nanotechnol* 2011; **6**; 147-150.
13. K. F. Mak, K. He, C. Lee, G. H. Lee, J. Hone, T. F. Heinz *et al.* Tightly bound trions in monolayer $MoS_2$. *Nat Mater* 2013; **12**; 207-211.
14. J. S. Ross, S. Wu, H. Yu, N. J. Ghimire, A. M. Jones, G. Aivazian *et al.* Electrical control of neutral and charged excitons in a monolayer semiconductor. *Nat Commun* 2013; **4**; 1474.
15. A. S. Rodin, A. Carvalho, A. H. Castro Neto. Excitons in anisotropic two-dimensional semiconducting crystals. *Phys Rev B* 2014; **90**; 075429.
16. A. Castellanos-Gomez, L. Vicarelli, E. Prada, J. O. Island, K. L. Narasimha-Acharya, S. I. Blanter *et al.* Isolation and characterization of few-layer black phosphorus. *2D Materials* 2014; **1**; 025001.
17. X. Wang, A. M. Jones, K. L. Seyler, V. Tran, Y. Jia, H. Zhao *et al.* Highly anisotropic and robust excitons in monolayer black phosphorus. *arXiv:1411.1695v1* 2014.
18. A. Favron, E. Gaufrès, F. Fossard, P. L. Lévesque, A.-L. Phaneuf-L'Heureux, N. Y.-W. Tang *et al.* Exfoliating pristine black phosphorus down to the monolayer: Photo-oxidation and electronic confinement effects. *arXiv:1408.0345v2* 2014.
19. V. Liu, S. Fan. S4 : A free electromagnetic solver for layered periodic structures. *Comput Phys Commun* 2012; **183**; 2233-2244.
20. H. Li, Q. Zhang, C. C. R. Yap, B. K. Tay, T. H. T. Edwin, A. Olivier *et al.* From bulk to monolayer $MoS_2$: Evolution of raman scattering. *Adv Funct Mater* 2012; **22**; 1385-1390.

21. J. Yang, Z. Wang, F. Wang, R. Xu, J. Tao, S. Zhang et al. Atomically thin optical lenses and gratings. *arXiv:1411.6200* 2014.
22. L. Liang, J. Wang, W. Lin, B. G. Sumpter, V. Meunier, M. Pan. Electronic bandgap and edge reconstruction in phosphorene materials. *Nano Lett* 2014; **14**; 6400-6406.
23. A. Thilagam. Two-dimensional charged-exciton complexes. *Phys Rev B* 1997; **55**; 7804-7808.
24. R. W. Keyes. The electrical properties of black phosphorus. *Phys Rev* 1953; **92**; 580-584.
25. E. Wigner. On the interaction of electrons in metals. *Phys Rev* 1934; **46**; 1002-1011.
26. M. Yoshikawa, M. Kunzer, J. Wagner, H. Obloh, P. Schlotter, R. Schmidt et al. Band-gap renormalization and band filling in si-doped gan films studied by photoluminescence spectroscopy. *J Appl Phys* 1999; **86**; 4400-4402.
27. M. Xiaodong, I. B. Zotova, Y. J. Ding, H. Yang, G. J. Salamo. Observation of an anomalously large blueshift of the photoluminescence peak and evidence of band-gap renormalization in inp/inas/inp quantum wires. *Appl Phys Lett* 2001; **79**; 1091-1093.
28. Y. Miyauchi, M. Iwamura, S. Mouri, T. Kawazoe, M. Ohtsu, K. Matsuda. Brightening of excitons in carbon nanotubes on dimensionality modification. *Nat Photonics* 2013; **7**; 715-719.
29. M. Iwamura, N. Akizuki, Y. Miyauchi, S. Mouri, J. Shaver, Z. Gao et al. Nonlinear photoluminescence spectroscopy of carbon nanotubes with localized exciton states. *ACS Nano* 2014; **8**; 11254-11260.
30. H. Wang, C. Zhang, W. Chan, C. Manolatou, S. Tiwari, F. Rana. Radiative lifetimes of excitons and trions in monolayers of metal dichalcogenide $MoS_2$. *arXiv:1409.3996* 2014.
31. V. Huard, R. Cox, K. Saminadayar, A. Arnoult, S. Tatarenko. Bound states in optical absorption of semiconductor quantum wells containing a two-dimensional electron gas. *Phys Rev Lett* 2000; **84**; 187-190.
32. K. Kheng, R. T. Cox, M. Y. d' Aubigné, F. Bassani, K. Saminadayar, S. Tatarenko. Observation of negatively charged excitons in semiconductor quantum wells. *Phys Rev Lett* 1993; **71**; 1752-1755.
33. C. Mai, A. Barrette, Y. Yu, Y. G. Semenov, K. W. Kim, L. Cao et al. Many-body effects in valleytronics: Direct measurement of valley lifetimes in single-layer $MoS_2$. *Nano Lett* 2013; **14**; 202-206.
34. H. Shi, R. Yan, S. Bertolazzi, J. Brivio, B. Gao, A. Kis et al. Exciton dynamics in suspended monolayer and few-layer $MoS_2$ 2D crystals. *ACS Nano* 2012; **7**; 1072-1080.
35. O. Salehzadeh, N. H. Tran, X. Liu, I. Shih, Z. Mi. Exciton kinetics, quantum efficiency, and efficiency droop of monolayer $MoS_2$ light-emitting devices. *Nano Lett* 2014; **14**; 4125-4130.

## Author Contributions

Y. R. L. designed the project; J. Y. did the PL measurements, data analysis and TMD semiconductor sample preparation; R. J. X. conducted PL data fitting analysis, few-layer phosphorene sample preparation and part of the theoretical calculations; J. J. P. conducted the preparation of monolayer phosphorene and PSI measurements; Y. W. M. contributed to sample preparation; F. W. built the optical characterization setup; Z. W. and Z. F. Y performed the theoretical simulations for the OPL of phosphorene. All authors contributed to the manuscript.

## Supplementary information

Supplementary information for this article can be found on the Light: Science & Applications' website (http://www.nature.com/lsa/).

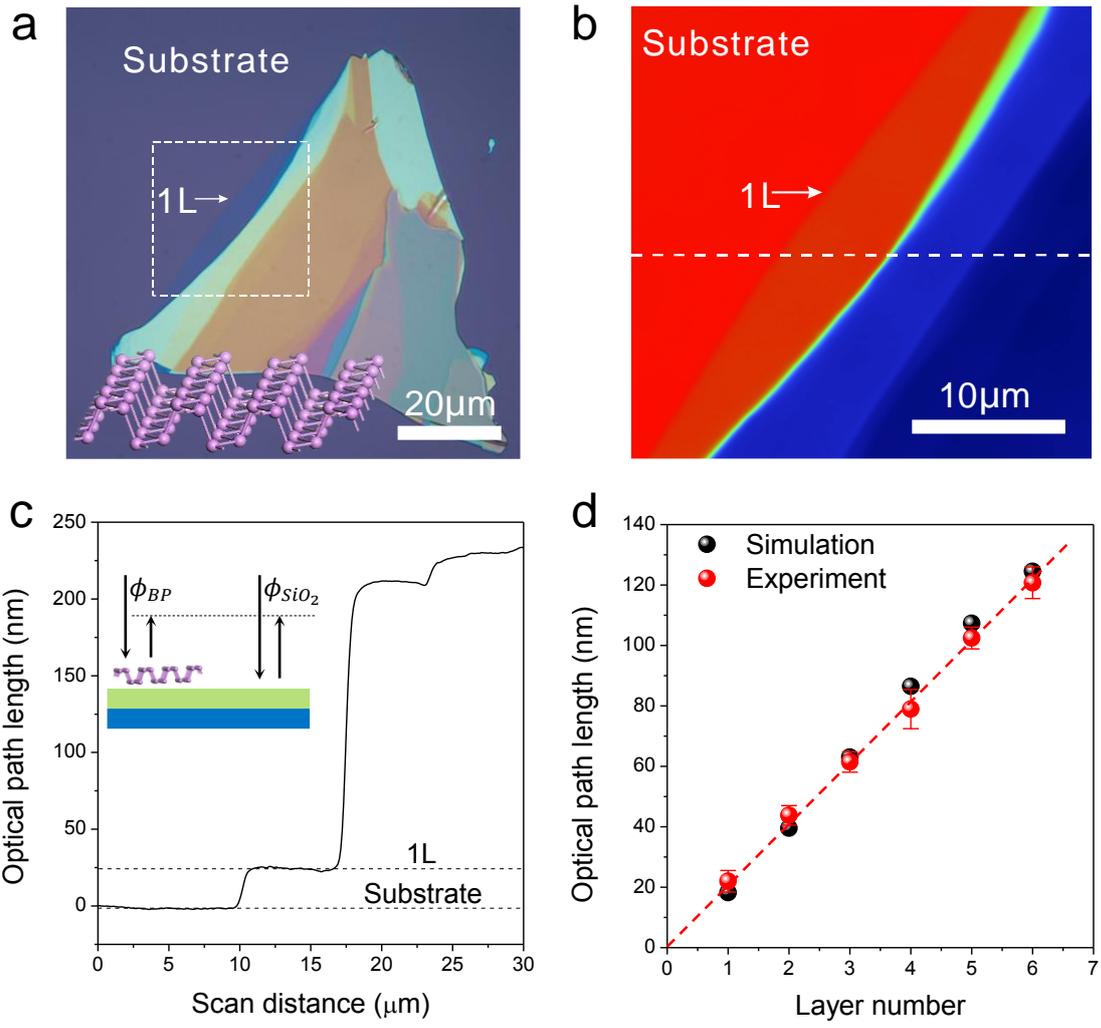

Figure 1

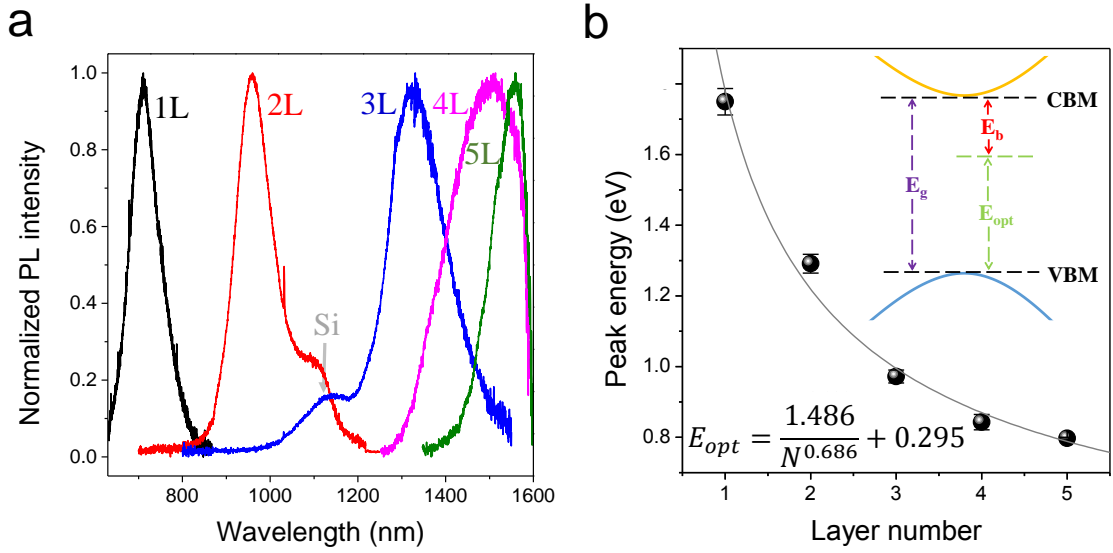

Figure 2

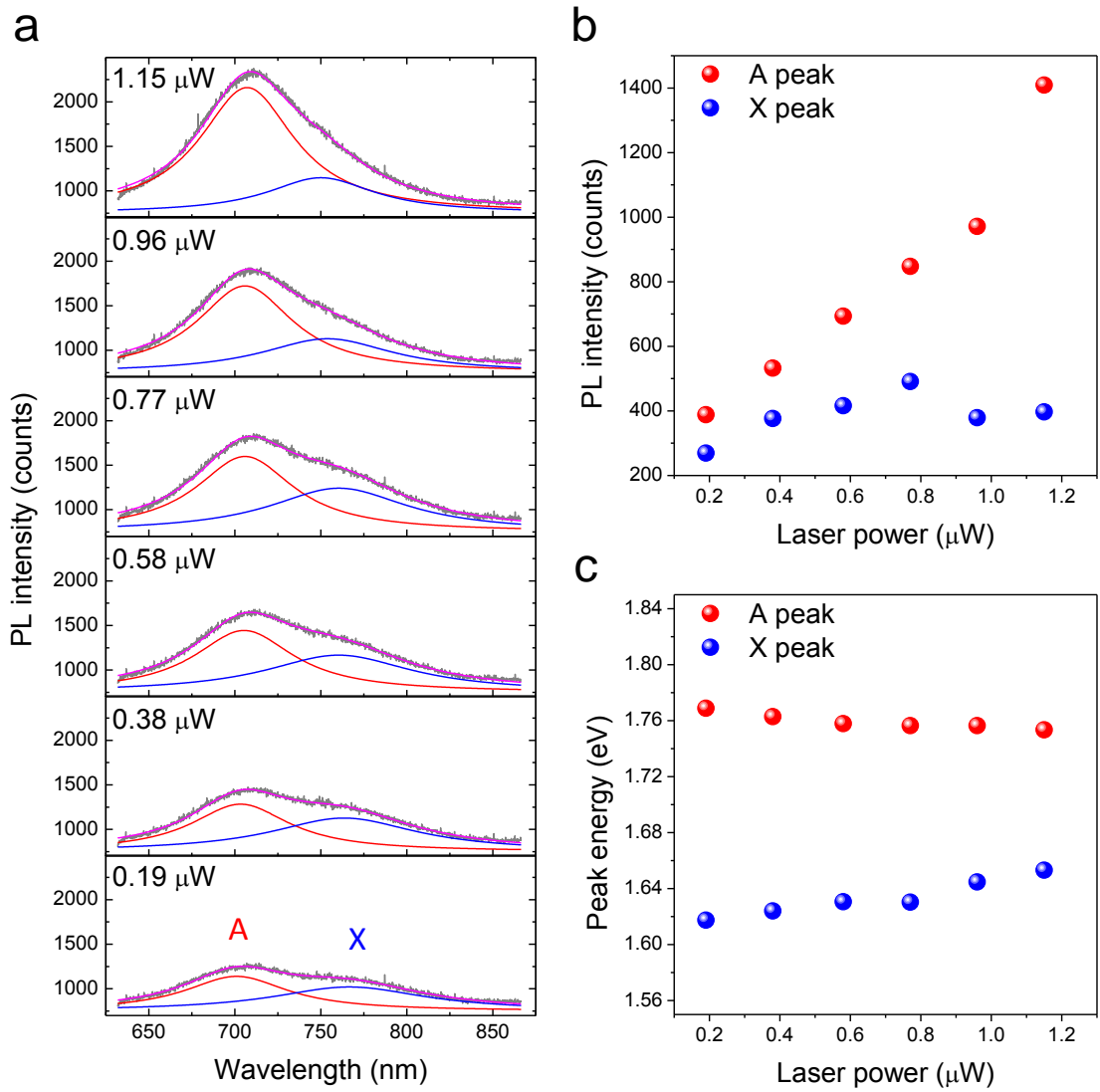

Figure 3

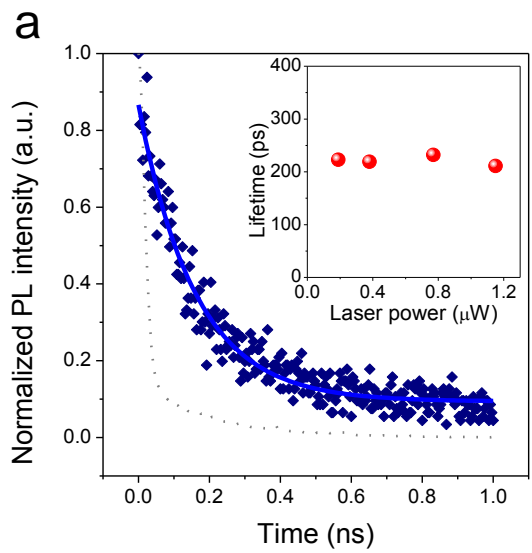 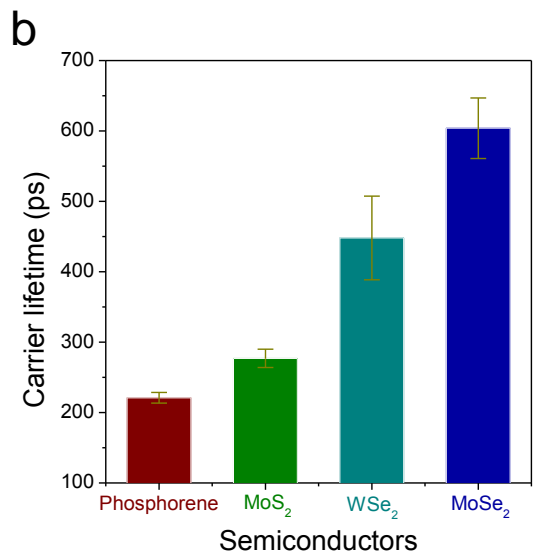

Figure 4

Supplementary Information for

# Optical Tuning of Exciton and Trion Emissions in Monolayer Phosphorene

Jiong Yang,[1,†] Renjing Xu,[1,†] Jiajie Pei, [1,4,†] Ye Win Myint,[1] Fan Wang,[2] Zhu Wang,[3] Shuang Zhang,[1] Zongfu Yu,[3] and Yuerui Lu[1*]

[1]Research School of Engineering, College of Engineering and Computer Science, the Australian National University, Canberra, ACT, 0200, Australia

[2]Department of Electronic Materials Engineering, Research School of Physics and Engineering, the Australian National University, Canberra, ACT, 0200, Australia

[3]Department of Electrical and Computer Engineering, University of Wisconsin, Madison, Wisconsin 53706, USA

[4]School of Mechanical Engineering, Beijing Institute of Technology, Beijing, 100081, China

[†] These authors contributed equally to this work

**\*** To whom correspondence should be addressed: Yuerui Lu (yuerui.lu@anu.edu.au)

**1. Characterization of another monolayer phosphorene sample**

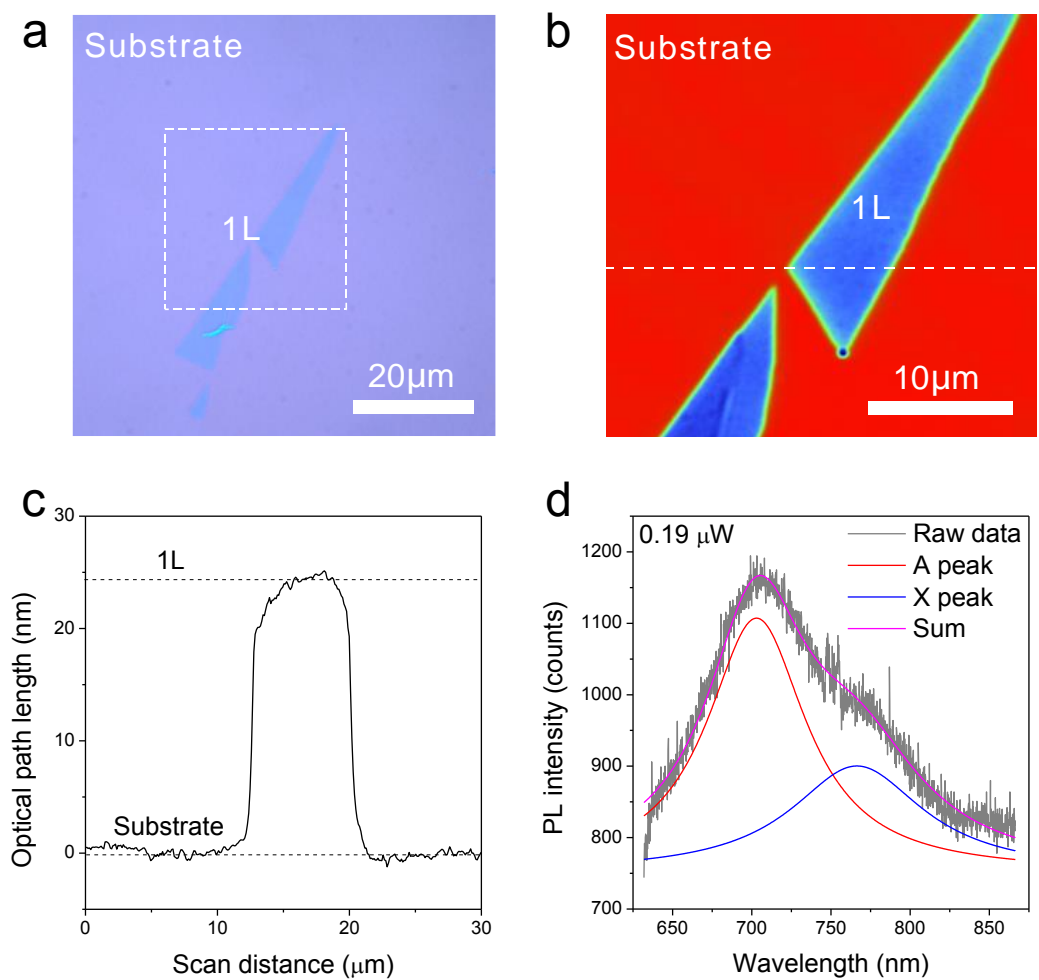

**Figure S1 | Characterization of another monolayer phosphorene. a,** Optical microscope image of monolayer phosphorene (labeled as "1L"). **b,** Phase shifting interferometry (PSI) image of the dash line box area indicated in (a). **c,** PSI measured optical path length (OPL) values along the dash line indicated in (b). **d,** Photoluminescence (PL) spectrum of the monolayer phosphorene sample in (a) measured with a pulse 522 nm green laser at a laser power of 0.19 μW (Solid grey curves are the measured raw data, solid red lines are the exciton (A) components, solid blue lines are the trion (X) components, and solid pink lines are the cumulative fitting results).

Regardless of the monolayer phosphorene sample presented in the main text, here we show another monolayer phosphorene sample and its identification and characterization. Figure S1a shows the optical microscope image of another monolayer phosphorene. It is further confirmed to be monolayer phosphorene with the PSI system. The OPL value of this monolayer sample is 24.5 nm, consistent with our OPL *v.s.* Layer number calibration curve indicated in Figure 1d. Figure S1d shows the measured PL spectrum with a 522 nm pulse laser at a laser power of 0.19 μW. Exciton and trion components were extracted to be at 702.9 nm and 766.3 nm, converted to energy as 1.76 eV and 1.62 eV, consistent with our aforementioned exciton and trion peak positions of monolayer phosphorene.

## 2. OPL values of mono- and few-layer phosphorene by PSI

Numerical calculations were carried out to calculate the OPL values for 2D materials with a thickness of 0.67 nm, shown as solid line in Figure S2a. It is noteworthy that the OPL values increase dramatically with the increase of materials' refractive indices. The OPL values for specific 2D semiconductors, including 0.67 nm $SiO_2$, 1L (0.34 nm) graphene, 1L (0.65 nm) phosphorene, 0.67 nm Si and 1L (0.67 nm) $MoS_2$ are presented in Figure S2a as well. In the numerical calculation, the refractive indices used for $MoS_2$[1], Si, phosphorene[2], graphene[3] and $SiO_2$ were $5.3 + 1.3i$, $4.15 + 0.04i$, $3.4$, $2.6 + 1.3i$ and $1.46$ respectively. Statistical data of OPL values measured with the PSI system and from the numerical calculations for graphene, phosphorene and $MoS_2$ samples from 1L to 4L are presented in Figure S2b and Figure S2c, respectively. Both the measured and calculated OPL values for these three semiconductors show a linear relationship with the layer number and they consist well with each other.

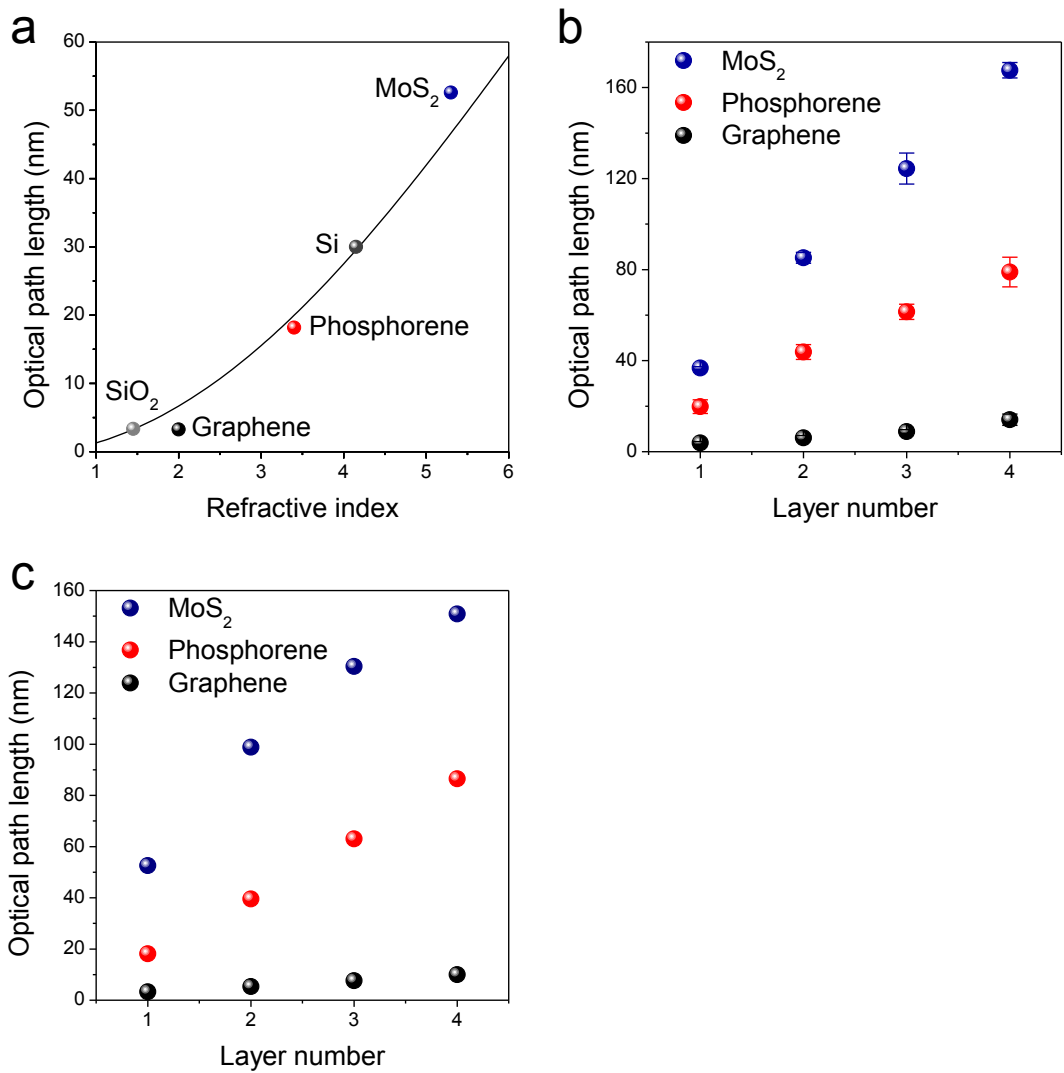

**Figure S2 | OPL values of mono- and few-layer phosphorene in comparison with other materials. a,** Simulated OPL values for light reflected from 2D material (0.67 nm in thickness) with different indices on a $SiO_2$ (275 nm)/Si substrate (solid line). The calculated OPL values of 0.67 nm $SiO_2$, 1L (0.34 nm) graphene, 1L (0.65 nm) phosphorene, 0.67 nm Si and 1L (0.67 nm) $MoS_2$ are represented by markers. **b,** Statistical OPL values of graphene, phosphorene and $MoS_2$ samples from 1L to 4L. For each layer number of these three semiconductors, at least five samples were characterized to get the statistical data, with error bar shown. **c,** Numerical calculation of the OPL values for graphene, phosphorene and $MoS_2$ samples from 1L to 4L.

## 3. Laser induced degradation of monolayer phosphorene

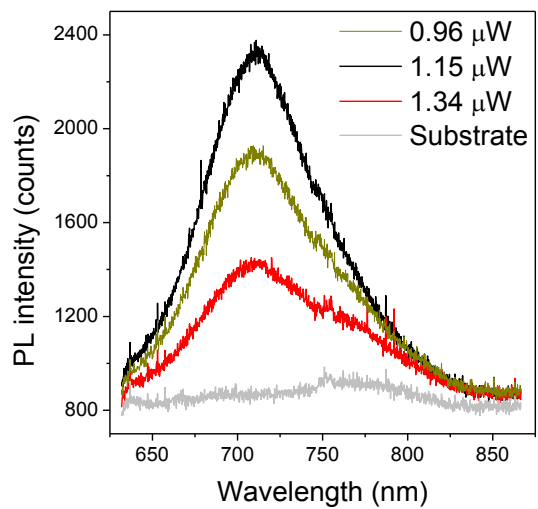

**Figure S3 | Laser induced degradation of monolayer phosphorene.** This sample is the one used for power-dependent PL measurements as shown in Figure 3. Black, dark yellow, red and grey curves represent monolayer phosphorene PL spectra with the laser power of 1.15, 0.96, 1.34 µW and PL spectrum on the $SiO_2$/Si substrate with a laser power of 1.15 µW.

During the power-dependent PL experiments, we found that high laser power ( > 1.15 µW) can damage our monolayer phosphorene samples, possibly by the laser induced oxidation. The PL spectra with the laser power of 0.96, 1.15 and 1.34 µW are presented in Figure S3, with the PL spectrum of $SiO_2$/Si substrate as reference. It is clear to see that the PL peak positions stay the same for PL spectra on the monolayer phosphorene sample. However, when the excitation laser power reaches 1.34 µW, the PL intensity begins to drop and cannot be recovered, due to a permanent damage to the monolayer phosphorene sample.

## 4. Temperature-dependent PL peak energies on 2L and 3L phosphorene

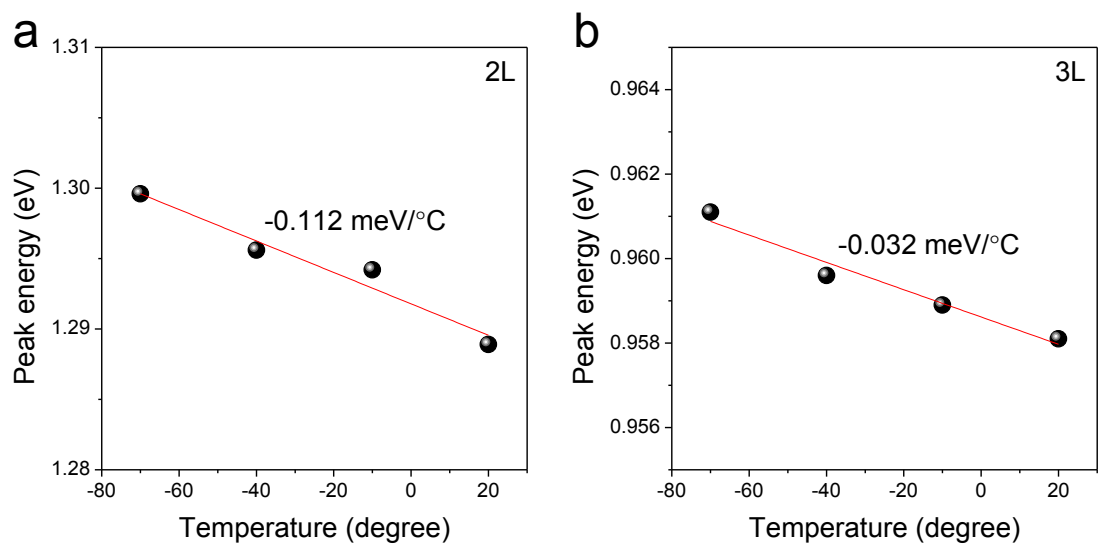

**Figure S4 | Temperature-dependent PL peak positions of 2L (a) and 3L (b) phosphorene samples.**

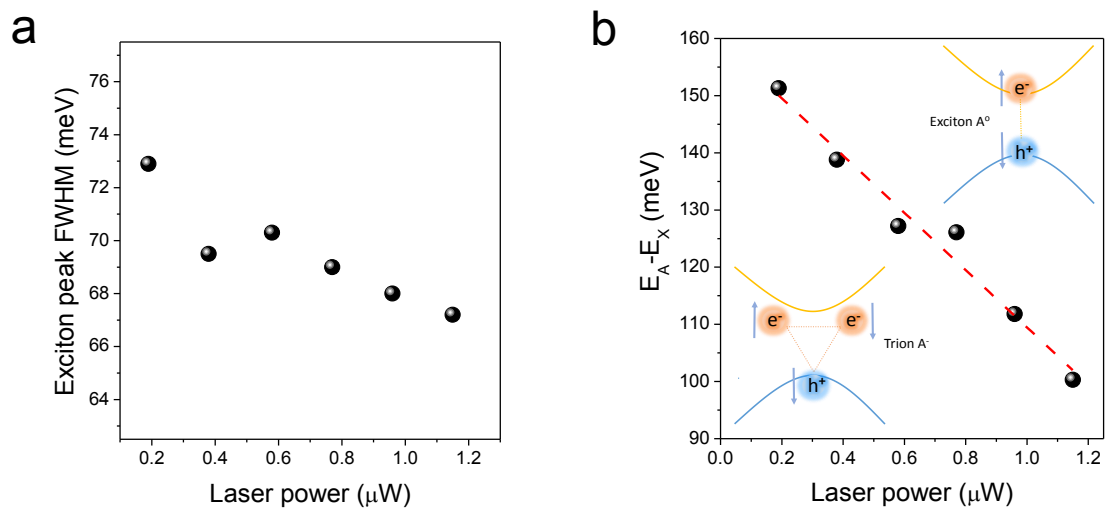

**Figure S5 | a**, Measured full width at half maximum (FWHM) of PL peak from exciton (A) in Figure 3a, as a function of the laser power. **b,** The difference in the PL peak energies of excitons (A) and trions (X) in Figure 3a, $E_A - E_X$ (symbols), as a function of laser excitation power. The red dashed line is a linear fit to the laser power-dependence. Inset: representations of a negatively charged trion (left bottom) and an exciton (top right).

## 5. Phase-shifting interferometry (PSI) working principle

PSI was used to investigate the surface topography based on analyzing the digitized interference data obtained during a well-controlled phase shift introduced by the Mirau interferometer[4]. The PSI system (Vecco NT9100) used in our experiments operates with a green LED source centered near 535 nm by a 10 nm band-pass filter[5]. The schematic of the PSI system is shown in Figure S6.

The working principle of the PSI system is as follows[6]. For simplicity, wave front phase will be used for analysis. The expressions for the reference and test wave-fronts in the phase shifting interferometer are:

$$w_r(x,y) = a_r(x,y)e^{i\phi_r(x,y)} \tag{S1}$$

$$w_t(x,y,t) = a_t(x,y)e^{i[\phi_t(x,y)+\delta(t)]} \tag{S2}$$

where $a_r(x,y)$ and $a_t(x,y)$ are the wavefront amplitudes, $\phi_r(x,y)$ and $\phi_t(x,y)$ are the corresponding wavefront phases, and $\delta(t)$ is a time-dependent phase shift introduced by the Mirau interferometer. $\delta(t)$ is the relative phase shift between the reference and the test beam. The interference pattern of these two beams is:

$$w_i(x,y,t) = a_r(x,y)e^{i\phi_r(x,y)} + a_t(x,y)e^{i[\phi_t(x,y)+\delta(t)]} \tag{S3}$$

The interference intensity pattern detected by the detector is:

$$I_i(x,y,t) = w_i^*(x,y,t) * w_i(x,y,t) = I'(x,y) + I''(x,y)\cos[\phi(x,y) + \delta(t)] \tag{S4}$$

where $I'(x,y) = a_r^2(x,y) + a_t^2(x,y)$ is the averaged intensity, $I''(x,y) = 2a_r(x,y) * a_t(x,y)$ is known as intensity modulation and $\phi(x,y)$ is the wavefront phase shift $\phi_r(x,y) - \phi_t(x,y)$.

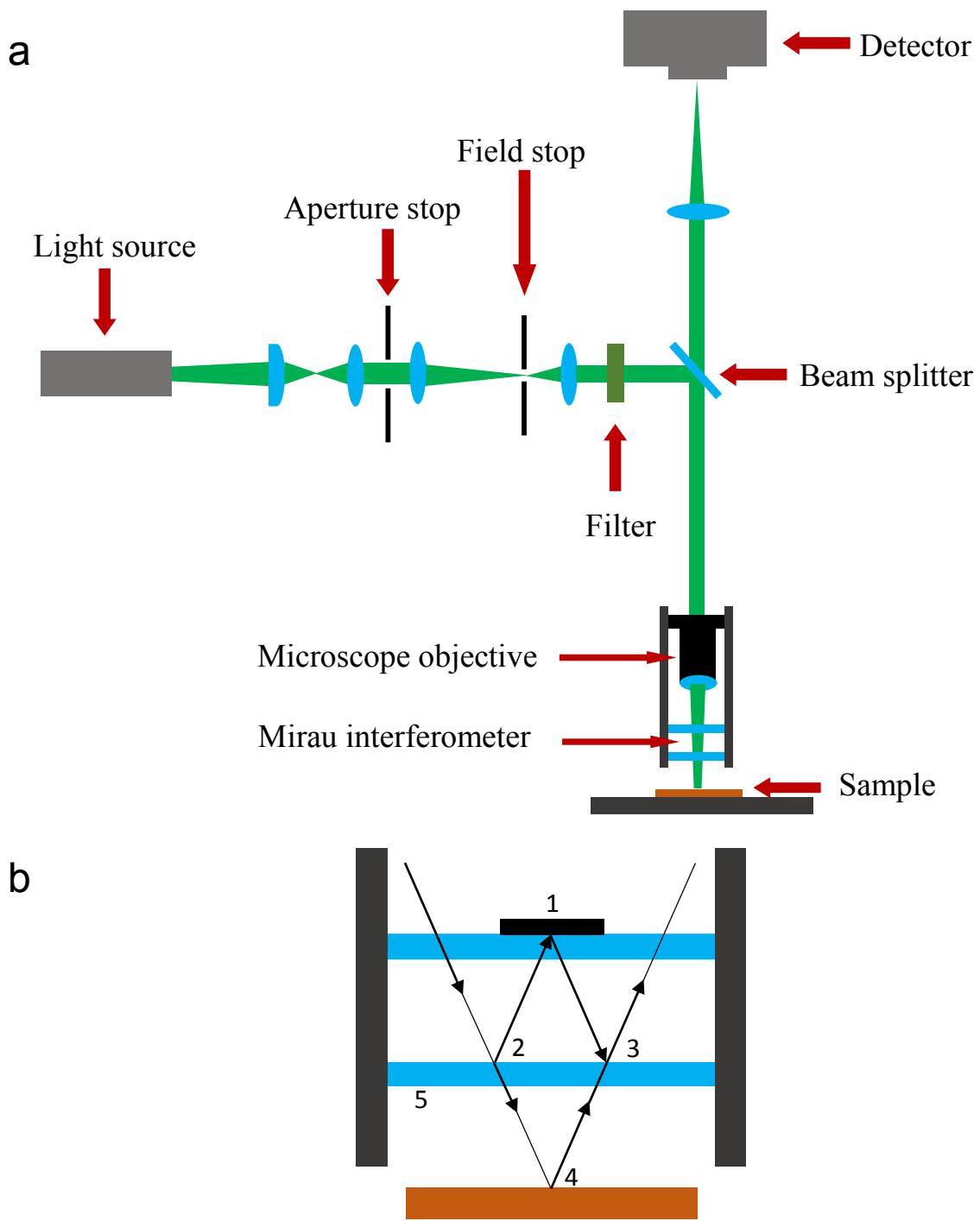

**Figure S6 | Schematic plot of the phase shifting interferometry (PSI) system. a,** Schematic plot of the PSI system. **b,** Zoomed view of the Mirau interferometer. 1. Reference mirror; 2. First reflection of the reference beam; 3. Third reflection of the reference beam; 4. Reflection of the test/objective beam; 5. Semi-transparent mirror. 2-1-3 represents the reference beam and 2-4-3 represents the test/objective beam.

From the above equation, a sinusoidally-varying intensity of the interferogram at a given measurement point as a function of $\delta(t)$ is shown below:

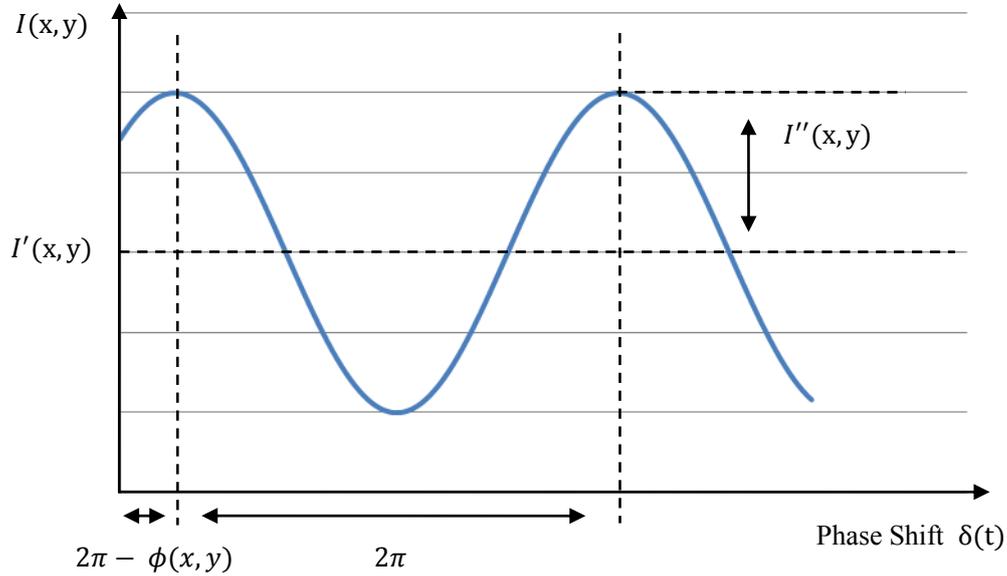

**Figure S7 | Variation of intensity with the reference phase at a point in an interferogram.** $I'(x,y)$ is the averaged intensity, $I''(x,y)$ is half of the peak-to-valley intensity modulation and $\phi(x,y)$ is the temporal phase shift of this sinusoidal variation.

$\delta(t)$ is introduced by the Mirau interferometer, which is shown in Figure S6. When the Mirau interferometer gradually moves toward the sample platform, the optical path length (OPL) of the test beam decreases while the OPL of the reference beam remains invariant.

The computational method of PSI is a four-step algorithm, which needs to acquire four separately recorded and digitalized interferograms of the measurement region. For each separate and sequential recorded interferograms, the phase shift difference is:

$$\delta(t_i) = 0, \frac{\pi}{2}, \pi, \frac{3\pi}{2}; \quad i = 1,2,3,4 \tag{S5}$$

Substituting these four values into the equation S4, leads to the following four equations describing the four measured intensity patterns of the interferogram:

$$I_1(x, y) = I'(x, y) + I''(x, y)\cos[\phi(x, y)] \tag{S6}$$

$$I_2(x, y) = I'(x, y) + I''(x, y)\cos[\phi(x, y) + \frac{\pi}{2}] \tag{S7}$$

$$I_3(x, y) = I'(x, y) + I''(x, y)\cos[\phi(x, y) + \pi] \tag{S8}$$

$$I_4(x, y) = I'(x, y) + I''(x, y)\cos[\phi(x, y) + \frac{3\pi}{2}] \tag{S9}$$

After the trigonometric identity, this yields:

$$I_1(x, y) = I'(x, y) + I''(x, y)\cos[\phi(x, y)] \tag{S10}$$

$$I_2(x, y) = I'(x, y) - I''(x, y)\sin[\phi(x, y)] \tag{S11}$$

$$I_3(x, y) = I'(x, y) - I''(x, y)\cos[\phi(x, y)] \tag{S12}$$

$$I_4(x, y) = I'(x, y) + I''(x, y)\sin[\phi(x, y)] \tag{S13}$$

The unknown variables $I'(x, y)$, $I''(x, y)$ and $\phi(x, y)$ can be solved by only using three of the four equations; but for computational convenience, four equations are used here. Subtracting equation S11 from equation S13, we have:

$$I_4(x, y) - I_2(x, y) = 2I''(x, y)\sin[\phi(x, y)] \tag{S14}$$

And subtract equation S12 from equation S10, we get:

$$I_1(x, y) - I_3(x, y) = 2I''(x, y)\cos[\phi(x, y)] \tag{S15}$$

Taking the ratio of equation S14 and equation S15, the intensity modulation $I''(x, y)$ will be eliminated as following:

$$\frac{I_4(x,y) - I_2(x,y)}{I_1(x,y) - I_3(x,y)} = \tan[\phi(x, y)] \tag{S16}$$

Rearranging equation S16 to get the wave-front phase shift term $\phi(x, y)$:

$$\phi(x, y) = \tan^{-1} \frac{I_4(x,y) - I_2(x,y)}{I_1(x,y) - I_3(x,y)} \tag{S17}$$

This equation is performed at each measurement point to acquire a map of the measured wave-front. Also, in PSI, the phase shift is transferred to the surface height or the optical path

difference (OPD):

$$h(x,y) = \frac{\lambda \phi(x,y)}{4\pi} \tag{S18}$$

$$OPD(x,y) = \frac{\lambda \phi(x,y)}{2\pi} \tag{S19}$$

Here, the OPL of the phosphorene flake $OPL_{BP}$ is calculated as:

$$OPL_{BP} = -(OPD_{BP} - OPD_{SiO_2}) = -\frac{\lambda}{2\pi}(\phi_{BP} - \phi_{SiO_2}) \tag{S20}$$

where $\lambda$ is the wavelength of the light source, $\phi_{BP}$ and $\phi_{SiO_2}$ are the measured phase shifts of the reflected light from the phosphorene flake and the SiO$_2$ substrate, respectively. In our experiments, $\phi_{SiO_2}$ was typically set to be zero, as shown in Figure 1c.

## 6. Calculations for the optical path length (OPL) of atomically thin 2D materials

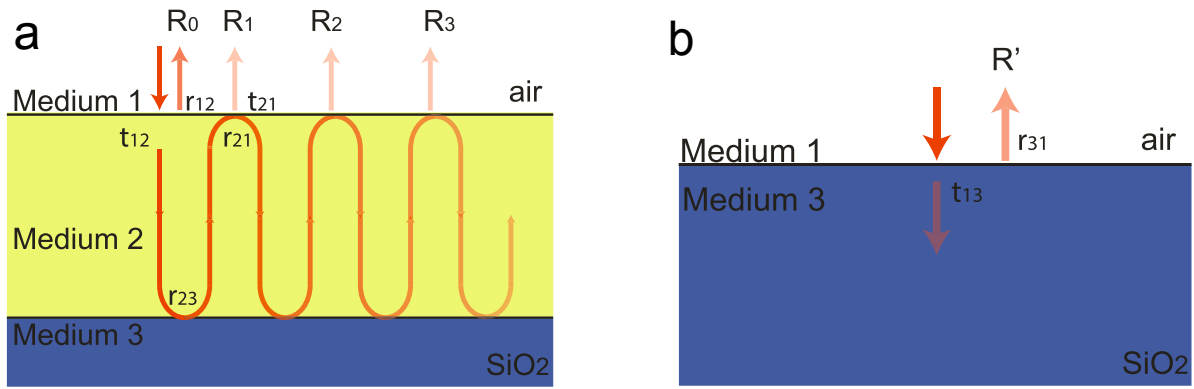

**Figure S8 | a**, Reflection of a three-layer structure. Medium 1 is air, Medium 2 is the 2D material and Medium 3 is an infinite SiO$_2$ substrate. **b**, The reference configuration. Light is incident from air into infinite SiO$_2$ substrate.

The incident light comes from the air resonates inside the 2D material. The total reflection is determined by the interference of all reflected beams $R_i$. To calculate the amplitude of the total reflection, we use $r_{ij}$ ($i,j=1,2,3$) to represent the reflection coefficients when light goes from medium *i* to medium *j*.

$$r_{ij} = \frac{n_i - n_j}{n_i + n_j} \tag{S21}$$

We use $t_{ij}$ (*i,j* =1,2,3) to represent the transmission from medium *i* to medium *j*

$$t_{ij} = \frac{2n_i}{n_i + n_j} \tag{S22}$$

where $n_i$, $n_j$ (*i,j* =1,2,3) is the refractive index of medium *i,j*. Assuming that the thickness of the 2D material is *d* and wave vector of incident light in air is $k_0$, we can calculate the reflection of each order,

$$R_0 = r_{12}$$

$$R_1 = t_{12} r_{23} t_{21} e^{i2k_0 nd}$$

$$R_2 = t_{12} r_{23} r_{21} r_{23} t_{21} (e^{i2k_0 nd})^2$$

$$R_3 = t_{12} r_{23} r_{21} r_{23} r_{21} r_{23} t_{21} (e^{i2k_0 nd})^3 \tag{S23}$$

where $2k_0 nd$ is the round trip propagation phase and *n* is the refractive index of the 2D material. Then the total reflected amplitude is the summation of all reflections, which is

$$\begin{aligned} R = R_0 + R_1 + R_2 + \\ = r_{12} + t_{12} r_{23} t_{21} e^{i2k_0 nd} \left[ 1 + r_{21} r_{23} e^{i2k_0 nd} + \left( r_{21} r_{23} e^{i2k_0 nd} \right)^2 + \cdots \right] \\ = r_{12} + \frac{t_{12} r_{23} t_{21} e^{i2k_0 nd}}{1 - r_{21} r_{23} e^{i2k_0 nd}} \\ = \frac{1-n}{1+n} + \frac{4n}{(1+n)^2} \frac{(n-1.46)}{(n+1.46)} e^{i2k_0 nd} \frac{1}{1 - \frac{(n-1)(n-1.46)}{(n+1)(n+1.46)} e^{i2k_0 nd}} \end{aligned} \tag{S24}$$

Here we used refractive indices of air and $SiO_2$ as 1 and 1.46, respectively.

The OPL was calculated by comparing the phase difference of the reflected light with and without the 2D material. Figure S8b shows the reference setup. Light is incident directly from air into infinite $SiO_2$ substrate. In this case the reflected amplitude is

$$R' = \frac{n_1 - n_3}{n_1 + n_3} \tag{S25}$$

So we get:

$$OPL = -\frac{\left(phase(R)-phase(R')\right)}{2\pi}\lambda \tag{S26}$$

where $\lambda$ is the wavelength of light. For phosphorene OPL calculations, we used the measured refractive index from bulk black phosphorus crystals ($n$ = 3.4)[2].

**7. Images and characterization of phosphorene flakes by PSI**

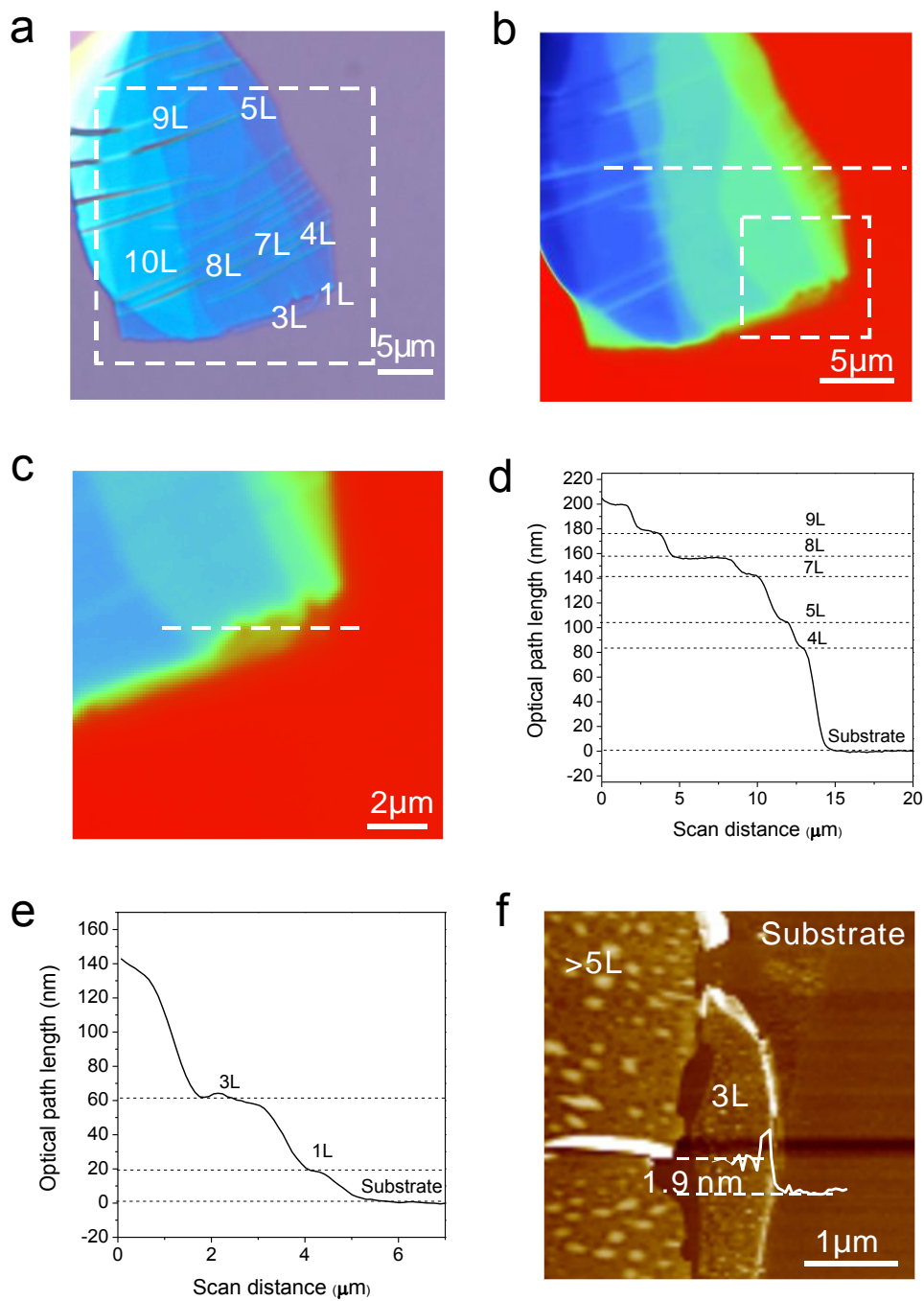

**Figure S9 | Images and characterization of exfoliated phosphorene. a,** Optical microscope image of a phosphorene flake containing multiple layers. **b,** PSI image of the phosphorene flake from the dash line box area indicated in (a). **c,** PSI image of the phosphorene flake from the dash line box area indicated in (b). **d** and **e** display the OPL measured by PSI versus position along the dash line in (b) and (c) respectively. **f,** AFM image of the 3L phosphorene flake.

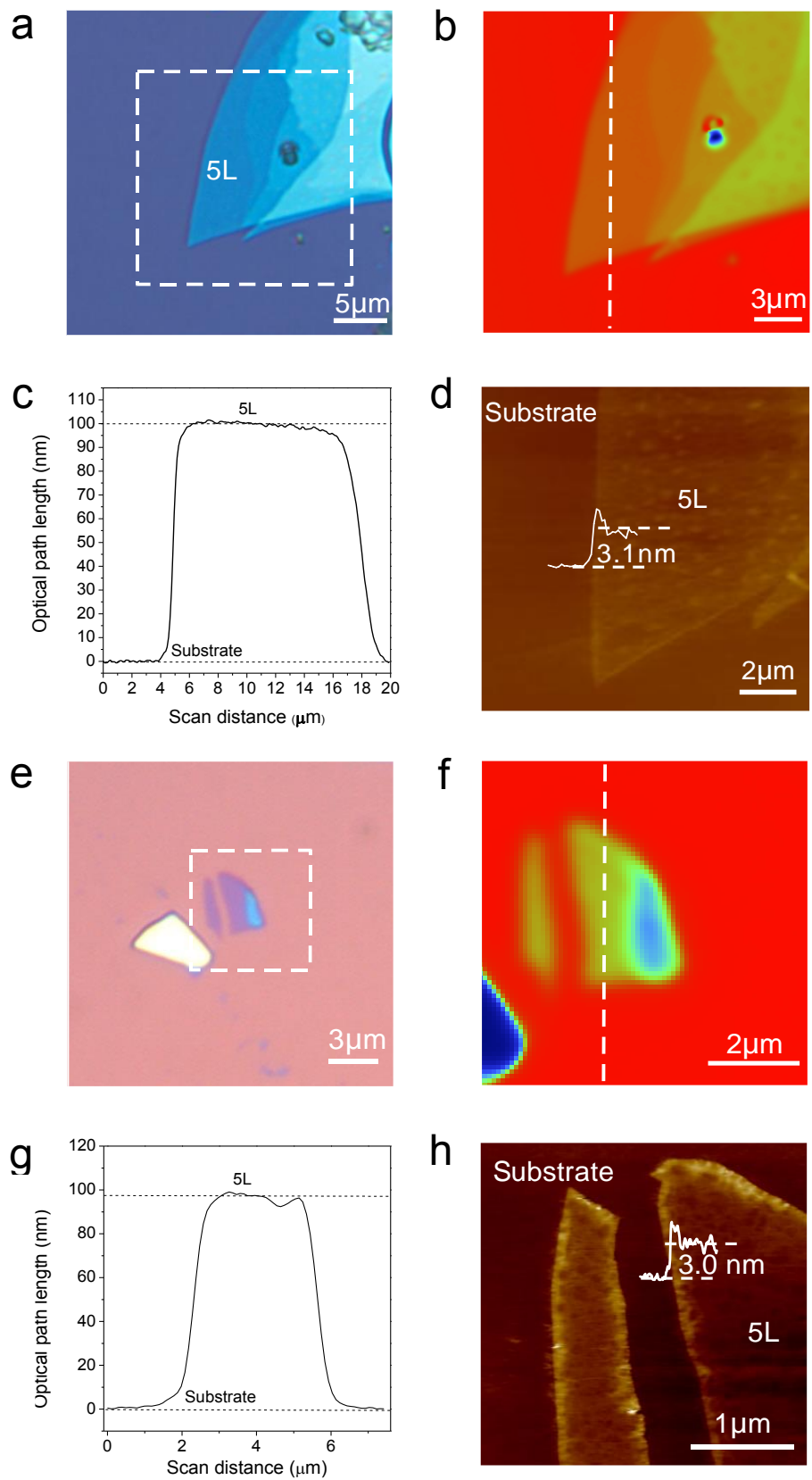

**Figure S10 | Images and characterization of exfoliated 5L phosphorene. a,** Optical microscope image of a 5L phosphorene flake. **b,** PSI image of the 5L phosphorene from the

dash line box area indicated in (a). **c,** OPL measured by PSI along the dash line in (b). **d,** AFM

image of 5L phosphorene. **e, f, g** and **h** display optical microscope image, PSI image (from the

dash line box area indicated in e), OPL (along the dash line in f) and AFM image of another 5L

phosphorene flake, respectively.

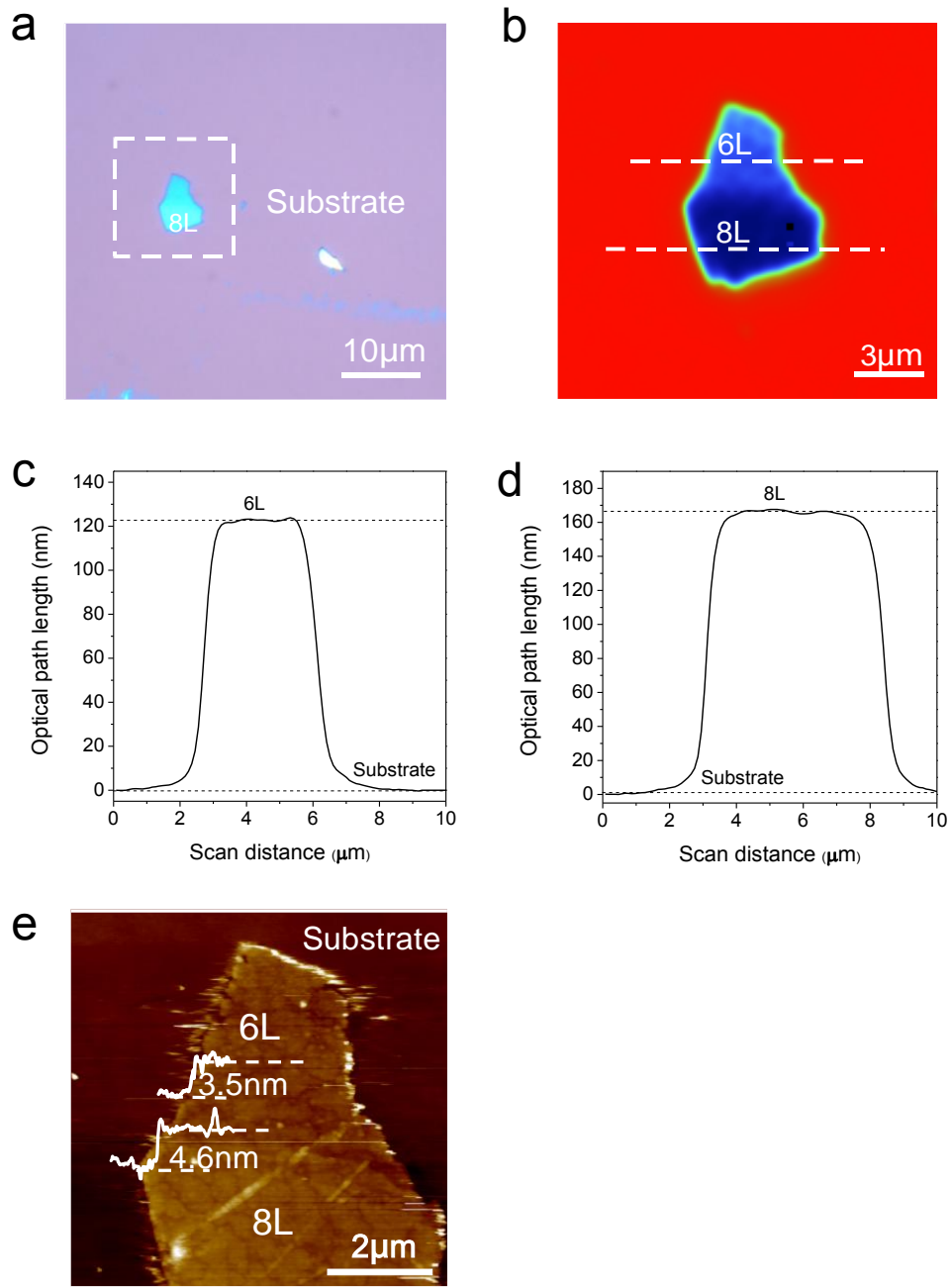

**Figure S11 | Images and characterization of exfoliated 6L and 8L phosphorene flakes. a,**

Optical microscope image of a phosphorene flake containing 6L and 8L. **b,** PSI image of the phosphorene flake from the dash line box area indicated in (a). **c** and **d** display OPL values measured by PSI from the 6L and 8L phosphorene along the dash lines in (b). **e,** AFM image of the 6L and 8L phosphorene.

## 8. PL measurement error analysis

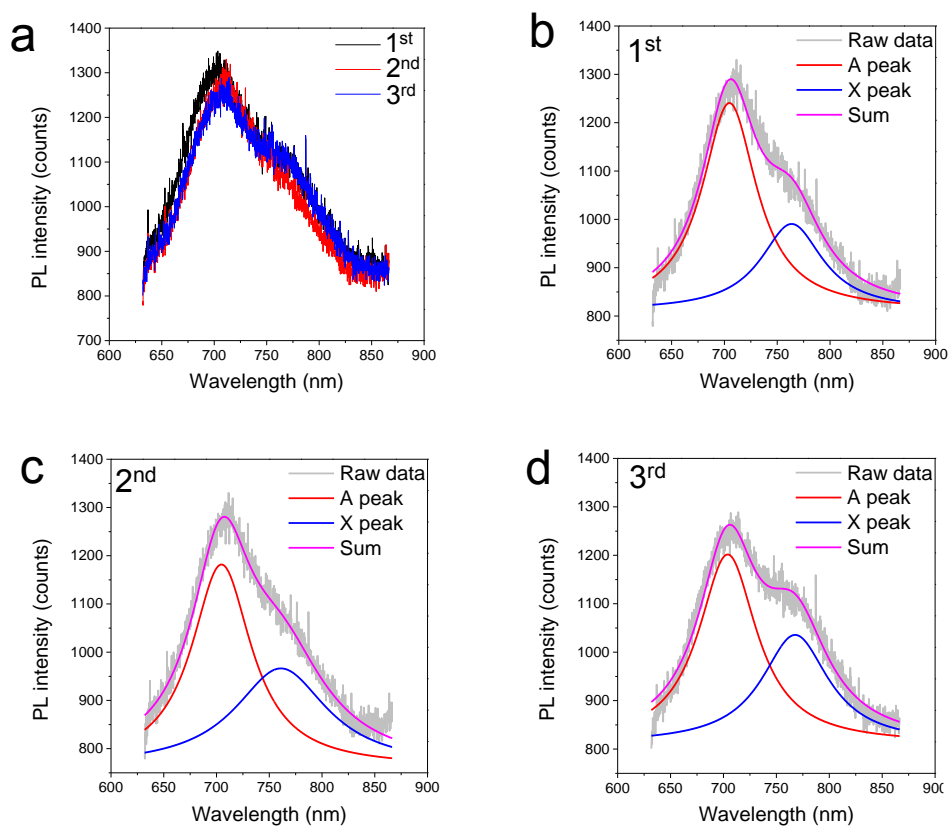

**Figure S12 | PL measurement error analysis. a,** PL spectra of the monolayer phosphorene sample measured with a pulse 522 nm green laser at a laser power of 0.19 μW, from three independent measurements. **b-d,** showing the PL spectra of the monolayer phosphorene sample from three measurements shown in (a), respectively, with Larentzian fitting (Solid grey curves are the measured raw data, solid red lines are the exciton (A) components, solid blue lines are the trion (X) components, and solid pink lines are the cumulative fitting results).

**Table S1: Detailed information of the measurement error from Figure S12.**

| Times | A peak position (eV) | A peak intensity (counts) | X peak position (eV) | X peak intensity (counts) |
|---|---|---|---|---|
| 1st | 1.772 | 452 | 1.619 | 204 |
| 2nd | 1.763 | 430 | 1.627 | 180 |
| 3rd | 1.765 | 391 | 1.618 | 224 |
| Average | 1.767 | 424 | 1.621 | 202 |
| Error@ | 0.005 | 31 | 0.005 | 22 |

Note: @ means standard deviation error.

9. **Optical injection estimation**

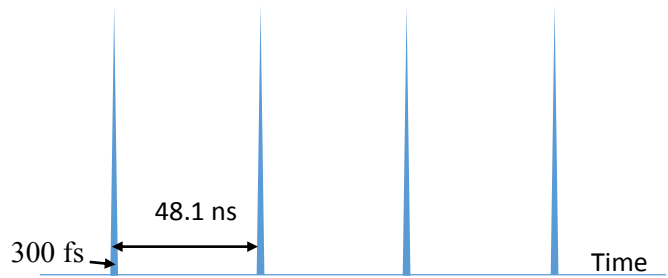

**Figure S13 | Schematic of the pulse laser used in our PL measurement.**

From Figure 3b, the intensity of X peak starts to saturate at the power of ~0.38 µW. So we will estimate the optical injection level at this power.

**Table S2. Parameters for the estimation of our optical injection.**

| | |
|---|---|
| Incident power (µW) | 0.38 |
| Laser wavelength (nm) | 522 |
| Pulse laser repetition rate (MHz) | 20.8 |
| Pulse width (fs) | 300 |
| Laser beam size (µm) | 0.6 |

The incident energy from each pulse of the laser is 0.38 µW*48.1 ns = 1.83 x $10^{-14}$ J. Based on published simulation result[7], the absorption of monolayer phosphorene could be ~5%. The absorbed photon number in each pulse is expected to be ~1.83 x $10^{-14}$ J/photon energy * 5% = 2.45 x $10^3$. With a pulse beam size of 0.6 µm and a monolayer thickness of ~0.7 nm, the volume of the injection region will be 1.98 x $10^{-16}$ $cm^3$. Since the pulse width (300 fs) is much smaller than the measured lifetime of excitons in monolayer phosphorene (~220 ps), we can assume that each of the absorbed photon would create one hole-electron pair. Based on our measured PL spectra data (Figure 3) at 0.38 µW excitation, the integrated PL intensity from the exciton emission peak (red) is ~1.02 times of that from the trion emission peak (blue). Therefore the trion volume density right after each pulse could be estimated to be ~2.45 x $10^3$/(1.98 x $10^{-16}$ $cm^3$)*(1/(1+1.02)) = 6.13 x $10^{18}$ $cm^{-3}$.

Next we will estimate the initial doping level of the monolayer phosphorene arising from dopants, which is assumed to be similar to the doping level of bulk black phosphorous crystals. The doping level of bulk black phosphorous crystal can be extracted from the published phosphorene field-effect transistor (FET) transport data by Ye et al[8]. Please note that both our crystal and Ye's crystal are from the same company (Smart Element).

**Table S3. Transport parameters of a phosphorene FET by Ye et al[8].**

| Hole mobility $\mu$ ($cm^2$/Vs) | 286 |
|---|---|
| Current $I$ at gate = 0 V (A/cm) | 1.1 |
| Thickness $t$ (nm) | 5 |
| Channel length $L$ (µm) | 1 |
| Source drain bias $V_{ds}$ (V) | 2 |

The velocity of holes in p-type phosphorene FET is:

$$v = \mu E = \frac{\mu V_{ds}}{L} = 5720000 \text{ cm/s}$$

where $\mu$ is the mobility of holes and $E$ is the electric field across the source and drain.

Then the initial doping level in the phosphorene FET could be extracted to be:

$$n = \frac{J}{ev} = \frac{I/t}{ev} = 2.4 \times 10^{18} \text{ cm}^{-3}$$

Where *J* is the current density and *e* is the elementary charge.

Therefore, the initial doping level in our monolayer phosphorene is estimated to be ~$2.4 \times 10^{18}$ cm$^{-3}$, which is comparable to the trion density of $6.13 \times 10^{18}$ cm$^{-3}$ estimated from our optical injection.

**References**

1. C.-C. Shen, Y.-T. Hsu, L.-J. Li, H.-L. Liu. Charge dynamics and electronic structures of monolayer MoS$_2$ films grown by chemical vapor deposition. *Appl Phys Express* 2013; **6**; 125801.
2. T. Nagahama, M. Kobayashi, Y. Akahama, S. Endo, S.-i. Narita. Optical determination of dielectric constant in black phosphorus. *J Phys Soc Jpn* 1985; **54**; 2096-2099.
3. P. Blake, E. W. Hill, A. H. Castro Neto, K. S. Novoselov, D. Jiang, R. Yang *et al.* Making graphene visible. *Appl Phys Lett* 2007; **91**; 063124.
4. R. Leach. *Optical measurement of surface topography*. Springer Berlin Heidelberg: **2011**; Vol. 8.
5. D. K. Venkatachalam, P. Parkinson, S. Ruffell, R. G. Elliman. Rapid, substrate-independent thickness determination of large area graphene layers. *Appl Phys Lett* 2011; **99**; 234106.
6. H. Schreiber, J. H. Bruning. Phase shifting interferometry. In *Optical shop testing*, John Wiley & Sons, Inc.: 2006; pp 547-666.
7. V. Tran, R. Soklaski, Y. Liang, L. Yang. Layer-controlled band gap and anisotropic excitons in few-layer black phosphorus. *Phys Rev B* 2014; **89**; 235319.
8. H. Liu, A. T. Neal, Z. Zhu, Z. Luo, X. Xu, D. Tománek *et al.* Phosphorene: An unexplored 2D semiconductor with a high hole mobility. *ACS nano* 2014; **8**; 4033-4041.